\documentstyle[12pt]{article}

\begin{document}

\begin{flushright}
gr-qc/0107086\\
$~$ \\
July 2001
\end{flushright}

\begin{centering}
\bigskip
{\leftskip=2in \rightskip=2in {\large \bf Space-time quantum solves
three experimental paradoxes}}\\
\bigskip
\bigskip
\bigskip
\bigskip
{\small {\bf Giovanni AMELINO-CAMELIA}}\\
\bigskip
{\small Dipartimento di Fisica, Universit\'{a} ``La Sapienza", P.le Moro 2,
I-00185 Roma, Italy}\\
\end{centering}

\vspace{1cm}
\begin{center}
{\bf ABSTRACT}
\end{center}

{\leftskip=0.6in \rightskip=0.6in

I show that a Planck-scale deformation of the relativistic dispersion
relation, which has been independently considered in the quantum-gravity
literature, can explain the surprising results
of three classes of experiments:
(1) observations of cosmic rays above the expected GZK limit,
(2) observations of multi-TeV photons from the BL Lac object
Markarian 501,
(3) studies of the longitudinal development
of the air showers produced by ultra-high-energy hadronic particles.
Experiments now in preparation,
such as the ones planned for the GLAST space telescope,
will provide an independent test of this solution of the
three experimental paradoxes.
}

\newpage
\baselineskip 12pt plus .5pt minus .5pt
\pagenumbering{arabic}
\pagestyle{plain}

Theoretical physics has been puzzling over the
structure of space-time at distance scales of the order of the
Planck length\footnote{$L_p \equiv \sqrt{\hbar G/c^3} \sim 1.6 {\cdot} 10^{-33}cm$,
where $\hbar$ is the reduced Planck constant,
$G$ is the gravitational constant and $c$ is the speed-of-light constant}
$L_p$ for several decades~\cite{stachisto,rovhisto,gacqm100}.
Unfortunately, there was no experimental counter-part for these
sizeable theoretical effort. All effects predicted by
Planck-scale theories are very small,
since they are strongly suppressed~\cite{rovhisto,gacqm100}
by the smallness of the Planck length, and this has kept
the Planck-length structure of space-time beyond the reach
of available experimental sensitivities.
Only over the last 15 years some ideas for experimental investigations
of this realm have started to
emerge (see, {\it e.g.},
Refs.~[4-16]),
relying on the remarkable sensitivities of
advanced experiments now in preparation.

At a time when it appeared to be rather exciting~\cite{rovhisto,gacqm100}
that some experiments could finally start exploring, in a few years,
the structure of space-time at the Planck scale,
it was recently
argued (see, {\it e.g.}, Refs.~\cite{kifu,aus,gactp})
that we might be already
witnessing the first manifestations of Planck-length physics,
since quantum-gravity models can provide solutions
for some
experimental paradoxes that presently confront the astrophysics
community: observed violations of the cosmic-ray GZK limit~\cite{gzk}
and observed violations of the analogous 10-TeV limit~\cite{aus,Coppi99}
that applies to photons from Markarian 501 (a BL Lac object at
a redshift of 0.034, {\it i.e.} $\sim 157$ Mpc).

I shall revisit these analyses of observations
of cosmic rays and Markarian-501 photons
and I shall then consider another independent
experimental paradox
which emerged from a very recent analysis~\cite{dedenko}
of data on the longitudinal development of the air showers produced by
ultra-high-energy hadrons.
Remarkably, I find that all three paradoxes are solved by the
same phenomenological model of
Planck-length physics, characterized by a deformed dispersion relation
(without free parameters!).
The preliminary evidence emerging from these three paradoxes
could amount to establishing the first
Planck-scale property of space-time.
Because of the profound significance of such a discovery
it is at present necessary to proceed very cautiously, and,
accordingly,
I shall also emphasize
the aspects of these three paradoxes
that still require further investigation.
The situation will be fully clarified, as I discuss
in the final part of this note, by forthcoming observations
by the GLAST space telescope~\cite{glastQG} which can
provide an independent and robust test of the scenario that
is being encouraged by the three experiments here analyzed.

Let me start briefly reviewing the three experimental paradoxes.
They all involve the kinematic rules for particle production
in a continuum classical space-time, but the relevant particle-production
processes are different and the energy scales involved are also
different:\\
{\bf Cosmic-ray paradox.}
Cosmic rays can interact with the Cosmic Microwave Background
Radiation (CMBR), producing pions. Taking into account
the typical energy of CMBR photons, and assuming the validity
of the kinematic rules for the production of particles
in our present, classical and continuous, description of space-time
(conventional relativistic kinematics),
one finds that these interactions should lead to
an upper limit $E < 5 {\cdot} 10^{19}$eV, the GZK limit~\cite{gzk},
on the energy of observed cosmic rays.
Essentially, cosmic rays emitted with energies in excess
of the GZK limit should loose energy on the
way to Earth by producing pions,
and, as a result, should still satisfy the GZK limit
when detected in our observatories.
Instead, several cosmic-rays
above the GZK limit (with energies as high as $3 {\cdot} 10^{20}$eV)
have been observed~\cite{AgaWat}.\\
{\bf Markarian-501 paradox.} Just in the same way in which one obtains
the GZK limit for cosmic rays, one also obtains a limit
on the maximum energy of photons that can reach us from distant sources.
The relevant process for establishing this limit
is pair-production absorption of high-energy
photons due to interactions with the Far Infrared Background Radiation
(FIBR).
For the high-energy photons emitted by Markarian 501,
FIBR absorption should~\cite{aus,Coppi99} become efficient around $10$TeV.
Markarian-501 photons with higher energies should
collide with FIBR photons, disappearing into an electron-positron pair,
and should therefore not be able to reach our observatories.
Instead, Markarian-501 photons with energies as high
as $24$TeV have been observed~\cite{Aharonian99}.\\
{\bf Pion-stability paradox.}
The kinematic rules for the production of particles
also govern the structure of the air showers produced by
high-energy particles. In particular, they allow to predict
some features of the longitudinal development of the showers,
such as the probability distribution of the maximum depth
of the showers. Experimental data
on the longitudinal development of the air showers
produced by ultra-high-energy hadronic primaries
appear to be in disagreement~\cite{dedenko} with these predictions.
The analysis reported in Ref.~\cite{dedenko}
suggests that the observed longitudinal development of the air showers
could be explained by assuming that ultra-high-energy neutral pions are
much more stable than low-energy ones, as if, at ultra-high energies,
the available phase space
for decay in two photons was becoming smaller
(perhaps, at some energy, even vanishingly small~\cite{dedenko})
than the one predicted by
conventional relativistic kinematics.

As announced, I intend to show that these paradoxes can be solved
by adopting a deformation of the standard relativistic
dispersion relation $E^2 = p^2 + m^2$.
In the quantum-gravity literature there has been discussion of various
mechanisms for the emergence of deformed dispersion relations.
The most radically new of these scenarios is the one~\cite{dsr} in which
a deformed dispersion relation is assumed to emerge as a reflection
of the deformed symmetries of a quantum version\footnote{Of course,
quantum versions of Minkowski space-time
usually do not enjoy classical symmetries.
In fact one of the schemes considered in Ref.~\cite{dsr}
turned out to be connected with the $\kappa$-Minkowski
noncommutative space-time~\cite{majrue,kpoinann}.}
of (quasi-)flat space-time.
Alternatively, deformed dispersion relations can also
emerge as a property\footnote{Space-time foam could play the role of
a dispersion-inducing environment~\cite{grbgac,garaytest,gampul}.}
of the space-time foam background~\cite{wheelerFOAM},
as illustrated
by the phenomenological model considered in Ref.~\cite{grbgac}
and by the analysis of
loop-quantum-gravity ``weave states"~\cite{ashtrovsmol}
reported in Ref.~\cite{gampul}.
In attempts to unify space-time physics with quantum mechanics
one can also encounter
deformed dispersion relations as a result of the presence
of more ordinary (no foam) backgrounds; for example, string theory
in certain magnetic-field-like backgrounds admits an
effective-theory description
in terms of a field theory in noncommutative geometry
with associated emergence of deformed dispersion relations
(see, {\it e.g.}, Ref.~\cite{sussNC}.)

From the analysis reported here below,
the careful reader will easily realize that the key ingredient
for the solution of the mentioned three paradoxes is a deformation
of the dispersion relation. It appears plausible that more than one
of the quantum-gravity schemes that motivate the analysis of
such deformations would provide solutions of the paradoxes.
However, I shall here focus on the space-time-foam
phenomenological scheme advocated in Ref.~\cite{grbgac},
based on the dispersion relation\footnote{The precise all-order function $f$
is not discussed in Ref.~\cite{grbgac}, but present and forthcoming
experiments are anyway only sensitive to the leading-order correction
to the presently-adopted dispersion relation $E^2 = p^2 + m^2$.}
\begin{equation}
E^2 = f(E,p;m;L_p) \simeq p^2 + m^2 - L_p E p^2  ~.
\label{disp1}
\end{equation}
In fact, my analysis will be facilitated by the simplicity of this
dispersion relation (for example, in other schemes~\cite{gampul,sussNC}
one should worry about a polarization dependence in the analysis
of processes involving photons). Moreover, the fact that (\ref{disp1}),
unlike other proposed deformations of the dispersion relation,
has no free parameters (but see closing remarks on alternatives
with different sign and magnitude of the deformation term)
renders particularly significant the fact that three independent
experimental paradoxes find a common solution in this scheme.

Let me start with the cosmic-ray paradox.
As first observed by Kifune~\cite{kifu},
the deformed dispersion relation (\ref{disp1})
would affect the kinematics of particle-production processes, such as
photopion production ($p + \gamma \rightarrow p + \pi$, with the
usual notation $p$, $\gamma$, $\pi$ to denote protons, photons and
pions respectively)
which, as mentioned, is relevant for the cosmic-ray paradox.
Combining (\ref{disp1}) with equations for the conservation
of energy and momentum one finds that a collision between
a proton of energy $E$ and a CMBR photon of (much smaller) energy $\epsilon$
can produce a pion (and a proton) only if
\begin{equation}
E > {2 m_p m_\pi + m_\pi^2 \over 4 \epsilon}
+ L_p {(2 m_p + m_\pi)^3 \, m_\pi^3 \over 256 \, \epsilon^4} \left( 1
- {m_p^2 + m_\pi^2 \over (m_p + m_\pi)^2} \right) ~,
\label{crthresh}
\end{equation}
where $m_p$ ($m_\pi$) is the proton (pion) mass.
The $L_p \rightarrow 0$ limit of this condition of course
describes the conventional photopion-production threshold.
In spite of the smallness of $L_p$ the correction term turns
out to be significant for the cosmic-ray paradox
(the magnitude of the correction term is suppressed by $L_p$
but is boosted by the large ratios $m_p/\epsilon$, $m_\pi/\epsilon$).
In fact, one finds~\cite{kifu,gactp} that,
according to (\ref{crthresh}),
even at $E \sim 3 {\cdot} 10^{20}$eV
photopion production on CMBR photons is still not possible,
providing an explanation for the fact that cosmic rays
of such high energies are being observed.

As shown in Refs.~\cite{aus,gactp},
the Markarian-501 paradox can be explained in a completely analogous
manner. Combining (\ref{disp1}) with the relevant
equations for the conservation
of energy and momentum one finds that
the process $\gamma + \gamma \rightarrow e^- + e^+$
is only possible if
\begin{equation}
{\cal E} > {m_e^2 \over \epsilon} + L_p { m_e^6 \over 8 \epsilon^4}  ~,
\label{pairthresh}
\end{equation}
where $m_e$ is the electron mass, ${\cal E}$ is the energy of the (hard)
photon emitted by Markarian 501, and $\epsilon$ denotes again the
energy of the (soft) background photon (here assumed to be a FIBR photon).
The $L_p \rightarrow 0$ limit of (\ref{pairthresh}) of course
describes the conventional pair-production threshold.
The $L_p$-dependent term again represents a significant
correction ($m_e/\epsilon \gg 1$).
Substituting for $\epsilon$ some typical energies of FIBR photons
($\sim 0.005$ eV)
one finds that the correction term is sufficient to
forbid electron-positron pair production
even well above $E \sim 20$TeV,
consistently with the observations
reported in Ref.~\cite{Aharonian99}.

The cosmic-ray paradox and the Markarian-501 paradox
involve different energy scales and different collision
processes,
but admit the same type of description (they require an increase
in the theory estimate of the threshold energy)
and, as shown, admit a common solution based on (\ref{disp1}).
The pion-stability paradox emerging from the analysis reported
in Ref.~\cite{dedenko} is of a different type, since it
involves a particle-decay process rather than a collision process.
While for the cosmic-ray and the Markarian-501 paradoxes solutions
based on the deformed dispersion relation (\ref{disp1}) had already
been discussed in the literature~\cite{kifu,aus,gactp},
previous studies of schemes leading to (\ref{disp1})
had not noticed the associated emergence of neutral-pion
increased stability,
and actually had not noticed any implications
for particle-decay processes.
This is the main technical/theory result reported in the present note,
and, remarkably, its phenomenology implications for pion decay into
photons are in agreement with the indication that
has emerged from the analysis reported in Ref.~\cite{dedenko}.
Let me focus the analysis of the implications of (\ref{disp1})
for particle decay directly on the example
relevant for the pion-stability paradox:
the process $\pi \rightarrow \gamma + \gamma$
(for other particle-decay processes one can of course proceed
in strict analogy).
My observation is based on the kinematical condition that
establishes a relation between
the energy $E_\pi$ of the incoming pion, the opening angle $\theta$
between the
outgoing photons and the energy $E_\gamma$ of one of the photons
(the energy $E_\gamma'$ of the second photon
is of course not independent; it is given by
the difference between the energy
of the pion and the energy of the first photon).
This relation is found, as usual, by combining the dispersion relation,
here assumed to be described by (\ref{disp1}),
with the equations for the conservation
of energy and momentum. One finds
\begin{eqnarray}
m_\pi^2 &\! = \!&
[2 E_\gamma E_\gamma' + L_p E_\pi E_\gamma E_\gamma']
[1 - \cos(\theta)]
+ 2 L_p E_\pi E_\gamma E_\gamma' \nonumber\cr
&\! = \!&
[2 E_\gamma (E_\pi - E_\gamma)
+ L_p E_\pi E_\gamma (E_\pi - E_\gamma)]
[1 - \cos(\theta)]
+ 2 L_p E_\pi E_\gamma (E_\pi - E_\gamma) ~.
\label{pithresh}
\end{eqnarray}
In the $L_p \rightarrow 0$ limit (the limit that corresponds
to our present classical picture of space-time)
this kinematical condition of course reproduces the corresponding
result for conventional relativistic kinematics.
The $L_p$-dependent correction term starts to be significant
at pion energies of order $(m_\pi^2/L_p)^{1/3}$.
When $E_\pi^3 > 2 m_\pi^2/L_p$
one finds that some of the values of $E_\gamma$ which correspond
to viable decay processes according to conventional relativistic
kinematics are no longer available to the decay process
(in particular, since $1 - \cos(\theta)$ is always positive,
one must exclude all values of $E_\gamma$
such that $m_\pi^2 - 2 L_p E_\pi E_\gamma (E_\pi - E_\gamma) < 0$).
As one easily sees from (\ref{pithresh}),
this reduction of the available phase space starts
rather quietly (only a very small reduction)
at $E_\pi \simeq (2 m_\pi^2/L_p)^{1/3} \sim 10^{15}$eV,
but gets stronger and stronger as the pion energy increases.
This picture of pion decay, with the associated depletion of
the number of photons produced by ultra-high-energy neutral pions,
would explain the puzzling experimental data
discussed in Ref.~\cite{dedenko}.

Having added the pion-stability paradox to the cosmic-ray and
Markaria-501 paradoxes, we now have
three paradoxes that
are solved by adopting the Planck-scale-deformed
dispersion relation (\ref{disp1}). In the literature one does not find
any other indication of departures from a classical space-time picture
and it is easy to check~\cite{grbgac,kifu}
that in the (huge number of) experiments that are
consistent with the classical dispersion relation $E^2 = p^2 + m^2$
the correction term introduced in (\ref{disp1}) is completely
negligible (in order to compensate for the smallness of $L_p$ the
physical context must involve unusually large hierarchies between
some relevant energy/length scales, such as the ratios $m_p/\epsilon$,
$m_e/\epsilon$
and $E_\pi/m_\pi$ encountered in the analysis of the
paradoxes).
Therefore, we seem to be confronted with the exciting perspective
of having to replace the classical space-time picture with a
new (quantum) picture involving the Planck length.
I should stress however that, while ordinarily the combined indications
of three independent experiments are considered to be sufficient
for drawing definitive conclusions, the three experiments on which
my analysis is based are still subject to some residual elements
of doubt, and some prudence may be appropriate.

The cosmic-ray paradox is well established, in the sense that there can be
no residual doubt concerning the fact that cosmic rays with energies beyond
the GZK limit are being detected. There are however some alternative
(not less speculative~\cite{kifu})
possible explanations of the cosmic-ray paradox,
which exploit the fact that we are unable to identify the astrophysical
sources of these cosmic rays and that the identification of ultra-high-energy
cosmic ray as protons
is still subject to (however small~\cite{dedenko}) margins of uncertainty.

With respect to the Markarian-501 paradox the residual elements of doubt
are complementary to the case of cosmic rays.
In fact, we have a clear identification of the observed particles as photons
and equally clear is the identification of Markarian 501
as the source. However, while measurements of the CMBR have
become more and more accurate over the years,
measurements of the FIBR have reached
a satisfactory level of accuracy only very recently (see, {\it e.g.},
Ref.~\cite{firbMEASURE}) and the robustness and interpretation
of these recent experimental results may still be subject to further scrutiny.
This is of course significant for establishing the Markarian-501
paradox, since the likelyhood that a Markarian-501
photon above threshold would reach
our detectors also depends on the (density of the) FIBR.

Concerning the robustness of the pion-stability paradox,
besides the need for more accurate data (some of the graphs that
support the analysis reported in Ref.~\cite{dedenko}
show data points with significant error bars),
a key reason of residual concern
resides, in this author's opinion, in the role that
quantum chromodynamics (QCD)
playes in the structure of
the air showers produced by hadronic particles.
The analysis reported in Ref.~\cite{dedenko}
appears to provide rather convincing evidence of the fact that
within the presently-favoured
phenomenological model of the relevant QCD processes
(a model which has been found to be reliable in other contexts)
the assumption of increased pion stability  at
ultra-high energies leads to improved
agreement with the data
on the longitudinal development
of the air showers. However, certain quantitative estimates based
on QCD are still rather challenging
(in spite of the fact that QCD has been well understood
conceptually and at the level of the formalism for several years)
and it appears to be reasonable to wonder whether one should also
explore the possibility of modifying
the presently-favoured phenomenological model of QCD processes,
without introducing an increase in pion stability.

In summary, the status of the three paradoxes is rather robust,
but each of them (to different degrees) is still not completely
immune from potential weaknesses. Perhaps a greater
level of confidence should be attributed to the
analysis here being reported by considering the consistency
of the combined indications of the three experiments.
In particular,
from a strictly phenomenological viewpoint one could also contemplate
deformed dispersion relations with the same structure of (\ref{disp1})
but with the opposite sign choice for the correction term
and/or with a deformation scale which is significantly
different from the Planck scale.
But for all three paradoxes the solutions
require the same sign choice\footnote{With the
opposite sign choice the two
threshold conditions here considered would go in the opposite direction,
predicting that the process is allowed at even lower energies than
in the conventional theory (in clear contrast with the experimental
information). As one can easily see by changing the sign in front
of $L_p$ in (\ref{pithresh}),
the opposite sign choice would also not predict the
increase of pion stability here discussed.},
so in order to assume that the evidence emerging from these paradoxes
is the result of the preliminary nature of the experimental data
one should assume that the independent inaccuracies
of these data have somehow conspired
to point all in the same direction.
Similarly, the evidence emerging from the three paradoxes also has
a significant level of internal consistency for what concerns the
deformation length scale. It is easy to see
that the requirement of explaining the three paradoxes
imposes that this length scale cannot be much smaller
than $L_p$ (the solution of the Markarian-501 paradox
is lost already by decreasing the deformation length scale
by a factor of 100 or so, while the solutions of the cosmic-ray
paradox and of the pion-stability paradox
have a few more orders of magnitude margin).
On the other hand the deformation length scale certainly cannot be much
larger than $L_p$ because otherwise a disagreement would emerge with data
at lower energies~\cite{grbgac}. Therefore the requirement of explaining
the three paradoxes, besides imposing a consistent sign choice,
also imposes that the deformation length scale
be within a few orders of magnitude of the Planck length,
just as one would expect in light of the quantum-space-time
arguments that support (\ref{disp1}).

While some prudence is certainly appropriate,
we are clearly confronted with growing experimental evidence in favour of the
exciting perspective of having to modify our present classical description
of the short-distance structure of space-time. The issue will be completely
settled within a few years by experiments such as the ones planned
for the GLAST space telescope. As discussed in detail
in Refs.~\cite{grbgac,billetal,glastQG} these experiments are sensitive to
the implications of (\ref{disp1}) for the structure of bursts
of gamma rays that we detect from distant astrophysical sources,
an effect that (since it does not involve particle-production processes)
is completely independent from the effects here analyzed in
association with (\ref{disp1}). The expected sensitivity levels
of GLAST (which even extend several orders of magnitude beyond
the Planck-length choice of the deformation length scale)
are such that (\ref{disp1}) will be either fully confirmed
or completely rejected~\cite{glastQG}.

If the GLAST verdict does confirm the growing evidence
that is emerging from the experimental paradoxes here considered,
theoretical physics will find itself in a situation that is
amusingly analogous to the one that was created, a century ago,
by the Michelson-Morley experiments (which can be described
as providing evidence in favour of the dispersion
relation $E^2=p^2$ for photons, and forced a revolution in the
description of space-time, abandoning the Galileo-Newton picture).
This will include the need to establish which of the space-time
pictures that support deformed dispersion relations is actually realized
in Nature. At present the fact that (\ref{disp1}), without any free parameter,
explains all observations appears to be significant. But even if
(\ref{disp1}) is indeed realized in Nature we will still have to
consider two main alternative scenarios that can lead to (\ref{disp1}).
The scenario on which I focused here, which, as mentioned,
is based~\cite{grbgac,billetal,kifu}
on some conjectured properties of space-time foam,
is strongly characterized by the fact that space-time foam
could have a preferred frame~\cite{grbgac,garaytest,gampul,mexqg},
just like the classical aspects of the geometry of the space-time
of our Universe are such that it is possible to identify a
preferred frame\footnote{At the fundamental level the theory
does not have a preferred frame, but of course the field distributions
that correspond to a given solution of the equations
of dynamics can be used to identify a preferred frame (a frame in which
these field distributions acquire a certain characteristic form.)}
(a convenient frame for most applications is the one
in which the CMBR has the simplest properties~\cite{kifu}).
The second scenario
that supports (\ref{disp1}) is based~\cite{dsr} on the
opposite assumption: the quantum features of space-time would not
have a preferred frame, in which case the role of $L_p$ in (\ref{disp1})
should be enforced as an observer-independent condition.
As mentioned, this second scenario leads~\cite{dsr} to a fundamental picture
of space-time that is noncommutative~\cite{majrue,kpoinann},
and accordingly Lorentz transformations between different
frames would be governed by a quantum-algebra version of
the Lorentz algebra that had
emerged~\cite{kpoinold,kpoinann}
as part of the
mathematical-physics programme of systematic studies of
quantum deformations of classical groups and algebras.

It is rather compelling that the simple first scenario, here considered,
manages to explain the three paradoxes, in spite of its highly constrained
structure (no free parameters). Still,
it will of course be interesting
to compare the paradoxes with other schemes leading
to deformed dispersion relations,
and particularly the mentioned scenario~\cite{dsr}
that supports (\ref{disp1})
as an observer-independent property of (noncommutative) space-time.
This more delicate (both technically,
because of the complexity of noncommutative
geometry, and phenomenologically)
analysis is postponed to future studies~\cite{gacinprep}.



\baselineskip 12pt plus .5pt minus .5pt

\vfil

\end{document}